\documentclass[a4paper,11pt]{article}
\usepackage[utf8]{inputenc}
\usepackage[margin=1in]{geometry}
\usepackage{hyperref}
\usepackage{siunitx}
\usepackage{xspace}
\usepackage{natbib}
\usepackage{graphicx}
\usepackage{authblk}
\usepackage[symbol*]{footmisc}

\usepackage{amsmath}
\usepackage[normalem]{ulem}
\usepackage{slashed}
\usepackage{booktabs}
\usepackage[pdftex,table]{xcolor}
\usepackage{xfrac}
\usepackage{mathtools}
\usepackage{empheq}
\usepackage{multirow}
\usepackage{amssymb}
\usepackage{url}
\usepackage{comment}
\usepackage{paralist}

\definecolor{RoyalBlue}{rgb}{0.25,.41,.88}
\definecolor{WildStrawberry}{HTML}{EE2967}

\title{Future of Artificial Intelligence for Science in Japan 2024 Community Report}
\date{}
\author[1,2]{Yoshitaka Itow\thanks{(\href{mailto:itow@icrr.u-tokyo.ac.jp}{\color{blue}itow@icrr.u-tokyo.ac.jp})}}
\author[3,6]{Jia Liu\thanks{(\href{mailto:jia.liu@ipmu.jp}{\color{blue}jia.liu@ipmu.jp})}}
\author[4]{Hirokazu Maesaka\thanks{(\href{mailto:maesaka@spring8.or.jp}{\color{blue}maesaka@spring8.or.jp})}}
\author[1]{Vinicius Mikuni\thanks{(\href{mailto:vmikuni@nagoya-u.jp}{\color{blue}vmikuni@nagoya-u.jp })}}
\author[3,6]{Nhat-Minh Nguyen\thanks{(\href{mailto:nhat.minh.nguyen@ipmu.jp}{\color{blue}nhat.minh.nguyen@ipmu.jp})}}
\author[1,5,6]{Hironao Miyatake\thanks{(\href{mailto:hironao.miyatake@nagoya-u.jp}{\color{blue}hironao.miyatake@nagoya-u.jp})}}
\author[1,7]{Atsushi J. Nishizawa\thanks{(\href{mailto:atsushi.nishizawa@iar.nagoya-u.ac.jp}{\color{blue}atsushi.nishizawa@iar.nagoya-u.ac.jp})}}
\author[3,6]{Patrick de Perio\thanks{(\href{mailto:pdeperio@ipmu.jp}{\color{blue}pdeperio@ipmu.jp})}}
\author[8,9]{Daniel Ratner\thanks{(\href{mailto:ratner@jlab.org}{\color{blue}ratner@jlab.org})}}
\author[10]{Kazuhiro Terao\thanks{(\href{mailto:kterao@slac.stanford.edu}{\color{blue}kterao@slac.stanford.edu})}}
\author[3,6]{Leander Thiele\thanks{(\href{mailto:leander.thiele@ipmu.jp}{\color{blue}leander.thiele@ipmu.jp})}}
\author[11,12]{Omar Alterkait}
\author[10]{Francois Drielsma}
\author[13]{Rocio Garcia}
\author[14,15]{Masako Iwasaki}
\author[16]{Ahsani Hafizhu Shali}
\author[1]{Federica Tarsitano}
\author[1]{Takahiro Terada}
\author[10]{Junjie Xia}

\affil[1]{Kobayashi-Maskawa Institute for the Origin of Particles and the Universe (KMI), Nagoya University, Aichi 464-8602, Japan}
\affil[2]{RCCN, Institute for Cosmic Ray Research, The University of Tokyo}
\affil[3]{Center for Data-Driven Discovery, Kavli IPMU (WPI), UTIAS, The University of Tokyo, Kashiwa, Chiba 277-8583, Japan}
\affil[4]{RIKEN SPring-8 Center, Hyogo, Japan}
\affil[5]{Institute for Advanced Research, Nagoya University, Aichi 464-8601, Japan}
\affil[6]{Kavli IPMU (WPI), UTIAS, The University of Tokyo, Kashiwa, Chiba 277-8583, Japan}
\affil[7]{Gifu Shotoku Gakuen University, Gifu, Japan}
\affil[8]{Thomas Jefferson National Accelerator Facility, Newport News, VA, 23606, USA}
\affil[9]{Old Dominion University, Norfolk, VA, 23529, USA}
\affil[10]{SLAC National Accelerator Laboratory, Menlo Park, CA, 94025, USA}
\affil[11]{Tufts University, 574 Boston Ave, Medford, MA 02155, USA}
\affil[12]{The NSF AI Institute for Fundamental Interactions, MIT, Cambridge, MA 02139, USA}
\affil[13]{High Energy Nuclear Physics Laboratory, Cluster for
Pioneering Research, RIKEN, 2-1 Hirosawa, Wako, Saitama 351-0198, Japan}
\affil[14]{Osaka Metropolitan University Graduate School of Science, Osaka, Japan}
\affil[15]{University of Osaka Research Center for Nuclear Physics (RCNP), Osaka, Japan}
\affil[16]{Research Center for Accelerator Technology, National Research and Innovation Agency (BRIN), Indonesia}

\begin{document}

\maketitle

\section{Executive Summary}

Artificial intelligence and machine learning (AI/ML) are becoming essential tools for current and next-generation physics research. Across accelerator physics, cosmology and astrophysics, and neutrino physics, experiments are producing larger, more complex, and more heterogeneous data sets. At the same time, progress is increasingly limited not only by statistical precision, but also by experimental uncertainties, computational cost, simulation fidelity, operational complexity, and the challenge of preserving and sharing knowledge across large collaborations.

This white paper summarizes scientific challenges and AI/ML research opportunities identified through the FAIRS Japan 2024 unconference process~\cite{fairs2024}. The discussion focuses on three major physics domains: accelerator physics, cosmology and astrophysics, and neutrino physics. Although each domain has distinct scientific goals and experimental constraints, several common technical themes emerge: high-dimensional reconstruction, fast and accurate simulation, uncertainty propagation, simulation-to-data mismatch, anomaly detection, real-time decision-making, and shared infrastructure.

In accelerator physics, AI/ML can support the design and operation of complex beam facilities. Important opportunities include surrogate modeling for expensive simulations, design of new facilities,  anomaly detection and monitoring, virtual diagnostics, and online system control with minimal human involvement and fast turnaround. These methods can improve experimental performance, reduce operational cost, and increase reliability while respecting strict safety and hardware constraints.

In cosmology and astrophysics, AI/ML can improve the extraction of fundamental physics parameters from large and heterogeneous observations of the Universe. Key opportunities include simulation-based inference for precise parameter determination, neural compression, field-level inference, fast cosmological simulations, anomaly detection of new phenomena, and reproducible data and model infrastructure. These methods can help future surveys control systematic uncertainties, extract more information from complex data, and improve robustness against model misspecification and the ``unknown unknowns''.

In neutrino physics, AI/ML can improve event reconstruction, accelerate and unify simulation, improve rare-event detection, and assist experimental design and operation. Neutrino measurements are made from indirect observations and are affected by complex detector response, interaction modeling, missing information, and limited calibration data. AI/ML methods can help reconstruct complex final states, reduce systematic uncertainties, identify rare signals, and optimize detector and beamline configurations toward physics sensitivity.

A central observation is that many AI/ML challenges are not confined to a single field. Shared developments in surrogate modeling, differentiable simulation, reconstruction, domain adaptation, uncertainty propagation, anomaly detection, real-time control, and knowledge infrastructure could benefit all three communities simultaneously. Common tools, benchmarks, model hubs, data standards, and knowledge bases would reduce duplication of effort, speed-up adoption of novel techniques, and lower the barrier to entry for new researchers.

Realizing this potential requires more than applying existing AI/ML tools to isolated problems. It requires coordinated investment in reusable software, shared data and simulation resources, uncertainty-aware methods, interpretable models, validation benchmarks, and collaboration structures that reward cross-domain development. With such coordination, AI/ML can become not only a collection of analysis techniques, but a foundation for more efficient, reproducible, and scientifically powerful physics research.


\section{Introduction}
Our aim is to guide scientific AI/ML research in the physics domains. The document summarizes key scientific challenges and potential AI/ML research opportunities to address them. The domain science areas we target in this document are those covered during the FAIRS Japan 2024 event---which include accelerator physics, cosmology and astrophysics, and neutrino physics. For each scientific domain we briefly explain the scientific goals and scope, the data structure used for scientific research, the core challenges to address, and the opportunities that AI/ML can provide as solutions. 

\subsection{Accelerator Physics}


\paragraph{Scientific Scope.}
Accelerator physics involves the design and operation of particle accelerators, powerful machines with applications across fundamental physics, photon science, medicine, and industry. Within high-energy physics, accelerators are primarily tasked with colliding particles at high energies to understand the inner structure of matter and how fundamental building blocks of the universe behave and interact. Photon science facilities such as synchrotron light sources and free-electron lasers are another major class, with applications spanning atomic-level analysis in physics, chemistry, biology, and medicine.

\paragraph{Data and Measurement Landscape.}
Given the rarity of some possible outcomes of particle interactions, millions of events per second are produced at experiments such as the Large Hadron Collider to create a large dataset consisting of multiple physics processes. Accelerator facilities also produce large volumes of data from beam diagnostics, detector signals, control systems, and simulations used to model accelerator components and beam dynamics. In photon science, area detectors can generate image data at up to $\approx$100,000 frames per second, requiring efficient online processing pipelines.

\paragraph{Core Challenges.}
On the accelerator side, major challenges include design of future facilities, online control and optimization of performance, prediction, detection, and recovery from operation anomalies, and data analysis, virtual diagnostics, and new capabilities for users. These challenges arise across the design and operation of accelerator facilities, where performance, reliability, cost, and safety must be optimized together. Beam simulations can also be a rate-limiting step in design and tuning. Additionally, reconstructing the full six-dimensional phase space distribution from limited diagnostics remains an open challenge.

\paragraph{Opportunity for AI/ML.}
Colliders have many use-cases for advanced AI methods from all aspects of the experimental workflow, from designing detectors and accelerators, calibrating instruments, simulating and reconstructing particle tracking and collisions, and interpreting particle interactions. AI methods can also benefit operations, improving accelerator performance, robustness, and capabilities. These challenges connect accelerator physics to several broader AI/ML themes in experimental science, including optimization, control, simulation, anomaly detection, and data analysis. A particularly promising direction is the development of surrogate models that emulate accelerator components or beam dynamics, which can serve as the basis for a digital twin of the entire facility---enabling fast optimization, predictive maintenance, and operator training without interrupting actual operations.

\subsection{Cosmology and Astrophysics}



\paragraph{Scientific Scope.}
Cosmology and astrophysics seek to infer fundamental physics properties from observations of the Universe. These observations are used to study the origin, composition, and evolution of the Universe, and to probe the physical processes that shape cosmic structure.

\paragraph{Data and Measurement Landscape.}
The relevant data include multi-wavelength observations of the Universe, including photometric imaging and spectroscopic redshift maps of galaxies, cosmic microwave background radiation maps, radio interferometry (e.g., gravitational wave) and time-domain observations (e.g., multi-messenger astronomy).

\paragraph{Core Challenges.}
Most inference tasks in this domain are ill-posed noisy inverse problems\footnote{Nonlinear gravitational evolution, astrophysical effects, selection and instrumental systematics all distort the underlying signal in nontrivial ways that are only partially invertible.}, for which Bayesian forward modeling is the optimal solution. Within this framework, one samples the model parameter space, simulates the data generation process, evaluates the likelihood, and repeats; this sampling procedure yields an estimate of the parameter posterior. The core challenges are either intractable likelihoods or---as the number of model parameter grows with data complexity---sampling efficiency.

\paragraph{Opportunity for AI/ML.}
ML methods like generative models blend well into several steps of the measurement and inference processes. In measurements and observations, ML algorithms improve source classification and outlier detection. In inference, neural networks can serve as data compression layers that reduce the dimensionality, and as model, likelihood, or posterior surrogates that either speed up the simulation and sampling processes or replace them entirely.

\subsection{Neutrino Physics}



\paragraph{Scientific Scope.}
Neutrino physics continues to advance rapidly along several scientific frontiers, seeking to uncover the fundamental properties of neutrinos, understand their role in the evolution of the universe, and exploit them as unique probes of astrophysical and geophysical phenomena. Long-baseline neutrino experiments will precisely probe neutrino oscillations to determine the neutrino mass ordering and search for charge-parity (CP) violation in the lepton sector, potentially providing an important clue to the origin of the matter–antimatter asymmetry in the universe. Neutrinos also provide unique windows into the physics of the Sun, supernovae, and high-energy cosmic accelerators, as well as the Earth's interior and the conditions of the early universe when it was only about one second old.

\paragraph{Data and Measurement Landscape.}
Neutrino experiments use a variety of detector technologies, including liquid argon time projection chambers (LArTPC), and large-volume optical detectors. Measurements are indirect, requiring reconstruction of neutrino properties from complex event signatures involving multiple particles and detector responses.

\paragraph{Core Challenges.}
Reconstructing neutrino interactions involves high-dimensional and complex event topologies, often with missing information and significant noise. Accurate and unified simulation of neutrino interactions and detector response is computationally demanding and subject to systematic uncertainties. In addition, rare signals must be identified against background, and experimental design and operation require navigating large parameter spaces with limited calibration and test-beam/prototype data.

\paragraph{Opportunity for AI/ML.}
These challenges motivate the use of AI/ML methods for reconstruction, ultra-rare event detection, simulation, and experimental design across current and future neutrino experiments. AI/ML methods can help classify and reconstruct complex final states, identify rare signals in large backgrounds, accelerate or improve simulation, and optimize detector designs toward scientific impact.

\section{Critical Science Challenges}

\subsection{Accelerator Physics}

Accelerators are among the largest, most-complex scientific facilities ever built. Designing, operating, and extracting science from these facilities all bring challenges that are well suited to AI methodologies. The accelerator-specific challenge is to improve the design and operation of complex beam facilities while respecting strict constraints on safety, reliability, cost, and performance.

\paragraph{Design.}
Designing new accelerator facilities can be a decade-long process involving hundreds of researchers. Increasingly, design is also computationally constrained, with a single high-fidelity simulation taking up to thousands of compute-hours. New methods are needed to both accelerate optimization of designs, as well as to extend the exploration of the space for design optimization. Recently a new approach is to tie together the entirety of the accelerator subsystems, extending to the detectors, into a single co-design process, bringing new opportunities but also new design challenges. Finally, traditional computational design is limited to optimization within predefined design configurations, with open challenges for high-level design of novel lattices, configurations, and accelerator components, and even exploration of novel beam dynamics and manipulations.

\paragraph{Operation.}
The largest accelerators today involve thousands of subsystems, monitor millions of process variables, and require teams of operators in control rooms around-the-clock. Improving performance, reducing cost, and increasing reliability are widespread challenges. Increasingly there is also need for more automation, both as accelerators gain complexity and also to enable technologies that can be used across facilities. Two accelerator-specific operational challenges have received particular attention: online optimization (``tuning'') to maximize performance, and anomaly (``fault'') prediction, detection, and recovery. Beyond achieving optimal performance, an equally important challenge is sustaining it autonomously over time. Beam conditions can drift due to environmental changes and component aging, and developing ML-based control algorithms capable of continuously maintaining optimal performance is an important open research problem.

\paragraph{Analysis and virtual diagnostics.}
The rapid increase in data collection at facilities has produced both challenges and opportunities for accelerators. First, the enormous amount of data threatens to overwhelm operations, creating a need for tools which can condense information to levels which can be processed by operators. On the other hand, there is an opportunity to develop new capabilities for the science users. For example, ``virtual diagnostics'' can provide information at rates, locations, or fidelities that would otherwise not be available, such as high-resolution information about beam or x-ray pulse characteristics.

\paragraph{Why it is challenging.}
Accelerator systems are high-dimensional, nonlinear, time-dependent, and safety constrained. Some calculations require many iterations and cannot simply be sped up by concurrent computing. Environmental drift, hidden dependence among subsystems, and hardware limitations further complicate optimization and operation. In many cases, the system must be improved without extensive experimental exploration, due to experimental cost or because unsafe parameter choices can damage components or interrupt operations.

\paragraph{Impact.}
Solving these challenges would improve accelerator performance, stability, and scientific output. Better machine performance and efficiency can reduce cost and enable beams with higher energy, intensity, and brightness. More robust dynamic and momentum aperture can produce higher luminosity at colliders and higher brightness at light sources. Better beam performance can improve downstream science, including high-brightness x-ray sources, intense neutrino beams, and high-luminosity collider operation.

\subsection{Cosmology and Astrophysics}

Data-driven cosmology and astrophysics face closely linked challenge areas that arise from extracting fundamental physics from noisy, incomplete, and high-dimensional observations of the sky. The central domain-specific challenge is to perform precise cosmological reconstruction and inference from large observational data sets while controlling theoretical, observational, and astrophysical systematics.

\paragraph{Observations and inference.}
Modern cosmological surveys provide images, maps, catalogs, time-domain data, and cross-correlations that contain information about the expansion history of the Universe, the growth of structure, dark matter, dark energy, and the physics of the early Universe. These data sets are large and heterogeneous, and useful information is distributed across multiple observables, wavelengths, and physical scales.

\paragraph{Systematics.}
As data volumes increase, statistical uncertainties are reduced, making systematic effects a dominant limitation. Important systematics include baryonic feedback, galaxy formation, photometric redshift uncertainties, selection functions, dust extinction, intrinsic alignments, source property assumptions, and mass modeling in strong lensing. Some of these effects can be modeled or calibrated, but others remain difficult to validate directly.

\paragraph{Why it is difficult.}
Cosmological inference is difficult because the underlying physical signal is distorted by nonlinear gravitational evolution, astrophysical effects, observational biases, and instrumental response. Large scales are relatively clean but information-poor, while small scales are information-rich but entangled with complex and uncertain physics. Observation is incomplete, simulations are approximate, and validation is often limited to simulations that may not fully capture real data.

\paragraph{Discovery in complex data.}
Future surveys will repeatedly scan large fractions of the sky and produce rich spatial, temporal, and multi-wavelength data. This creates opportunities to find rare transients, unusual lenses, unexpected source populations, poorly modeled systematics, or signatures of new physics. At the same time, rare signals must be separated from a vast background of ordinary astrophysical complexity and from instrumental or pipeline artifacts. Similar applications also benefit adjacent research fields such as gravitational wave research where signal candidates are rare and often background dominated.

\paragraph{Impact.}
Solving these challenges would allow future surveys to extract more information from galaxy clustering, weak lensing, cosmic microwave background measurements, and multi-wavelength observations. It would improve constraints on cosmic expansion, structure growth, primordial non-Gaussianity, parity violation, the nature of dark matter and dark energy, and neutrino mass. It would also make cosmological analyses more reproducible, interpretable, and robust against unknown unknowns.


\subsection{Neutrino Physics}

Neutrino physics faces several linked challenges: optimal event and particle reconstruction, accurate and unified simulation, robust treatment of systematic errors, anomaly detection for rare or transient processes, and optimal experimental design and operation. The domain-specific difficulty is that neutrino properties must be inferred indirectly from complex final states in large detectors, with missing information, large backgrounds, and limited calibration data.

\paragraph{Event reconstruction.}
A central challenge is to reconstruct neutrino properties, such as flavor, energy, direction, and interaction topology, from the final state particles observed in a detector. This reconstruction must work across a wide range of neutrino energies and detector technologies, including water and ice Cherenkov detectors, LArTPC, liquid scintillator detectors, and emulsion detectors.

\paragraph{Simulation and systematics.}
Accurate simulation is essential for neutrino physics because many measurements depend on detailed modeling of neutrino interactions, particle propagation, and detector response. Discrepancies between simulation and data lead to systematic biases. A major challenge is therefore to develop simulations that are accurate, unified, and fast enough to support large-scale analyses, while properly accounting for correlations between different physics and detector models.

\paragraph{Why it is difficult.}
Neutrino interactions span a broad range of event topologies, from simple final states to complex hadronic activity. Some particles are invisible or difficult to measure, and detector noise or limited resolution can degrade reconstruction. Neutrino--nucleus interactions, hadronic interactions, and detector response are also not perfectly understood, while calibration data are limited by in situ access and incomplete calibration coverage of the relevant energy and topology ranges.

\paragraph{Rare processes and experimental design.}
Rare and transient signals, such as supernova neutrinos or other low-rate processes, must be identified against detector noise and background. At the same time, experimental design and operation require exploring large parameter spaces of detector configurations, beamline choices, calibration strategies, and analysis methods. Current approaches often rely heavily on expert intuition and separate optimization of individual components.

\paragraph{Impact.}
Solving these challenges would improve sensitivity to neutrino oscillation parameters, CP violation, neutrino mass ordering, supernova neutrinos, and other rare processes. It would also improve cross-section measurements by reducing systematic uncertainties, make better use of beam uptime and detector exposure, and support more efficient design and operation of future neutrino experiments.


\subsection{Cross-Cutting Experimental Challenges}

Across experimental facilities, large amounts of data collected by detectors and instruments pose common challenges in event collection, computing resources, real-time processing, calibration, and interpretation. These challenges appear in different forms across accelerator facilities, cosmology surveys, and neutrino experiments. The shared challenge is that modern experiments produce high-dimensional, high-rate, heterogeneous data streams that must be reduced, reconstructed, calibrated, and interpreted without losing essential information.

\paragraph{Computing resources and fast reconstruction.}
Modern experiments increasingly operate at high rates and high occupancy, creating large demands on data taking, storage, and compute infrastructure. Event reconstruction, detector simulation, survey processing, beam diagnostics, and high-fidelity detector modeling all face the same tension between fidelity, speed, and scalability.

\paragraph{High-dimensional measurement space.}
Modern detectors and instruments consist of large numbers of readout channels, sensors, or detector components. Extracting information from raw readouts often requires dimensionality reduction or summary statistics, but this process is not always reversible or optimal. A shared challenge is to identify compact representations that preserve the information needed for reconstruction, inference, monitoring, and discovery.

\paragraph{Simulation-to-data mismatch and uncertainty propagation.}
Simulations are essential for interpreting experimental data, but even high-quality simulations require calibration to match observations. Machine learning models trained on simulations can be biased if the training domain does not match the deployment domain. At the same time, precision measurements require experimental, calibration, detector-response, and theoretical uncertainties to be propagated through complex analysis pipelines without requiring an infeasible number of full simulations.

\paragraph{Real-time processing, calibration, and monitoring.}
Many experiments must make fast decisions before all information can be stored. Trigger systems, online reconstruction, beam monitoring, detector control, and real-time calibration all require reliable low-latency processing. This is challenging because rejected data are often impossible to recover, detector conditions drift over time, and monitoring systems must distinguish real anomalies from benign fluctuations.

\paragraph{Anomaly detection and discovery.}
Identifying rare phenomena, detector failures, data-quality problems, or possible signs of new physics is a cross-cutting challenge. Since unexpected signals are not known in advance, anomaly detection is inherently ill-posed. Methods must be sensitive enough to identify rare or unexpected behavior while robust enough to avoid triggering false alarms, detector artifacts, and spurious discoveries.

\paragraph{Experimental design.} A variety of challenges require efficient search over complex parameter space, including design of next-generation facilities, optimal operation of facilities (accelerators, detectors, telescopes, etc.), and acquisition of new data for physics discovery. Constraints vary depending on the application, but include the high cost of data acquisition, high-dimensional design spaces, multi-fidelity inputs, and the need for uncertainty- and safety-aware exploration.

\paragraph{Impact.}
Solving these cross-cutting experimental challenges would reduce duplication of effort across domains and minimize the need for each community to ``reinvent the wheel.'' Shared developments in reconstruction, simulation, calibration, monitoring, and real-time processing could provide common tools, benchmarks, and knowledge bases. This would lower the barrier of entry for new experimentalists, improve reproducibility, and allow future facilities to extract maximal scientific value from increasingly complex data.


\section{Recommended AI/ML Research Directions}
\subsection{Accelerator Physics}

AI/ML solutions for accelerator physics focus on optimization, monitoring, virtual diagnostics, and fast decision-making. The common goal is to improve accelerator design and operation while respecting constraints from safety, reliability, cost, and limited facility time. The construction of surrogate models and digital twins that emulate accelerator components and beam dynamics is also a key research direction, providing fast and continuously updated virtual replicas of the physical machine.

\paragraph{Surrogate models and digital twins.}
Building data-driven surrogate models that emulate individual accelerator components or full beam dynamics at a fraction of the cost of conventional simulations is a key AI/ML research direction. Such surrogates can be integrated into a digital twin of the entire facility — a continuously updated virtual replica of the physical machine — enabling fast design optimization, online performance prediction, and operator training. Key research challenges include ensuring physical consistency of the surrogate outputs, handling distribution shift between simulation and real machine data, and keeping the digital twin synchronized with evolving machine conditions. These surrogates can be trained either through a static pipeline that separates data collection from model training, or through an iterative experimental design framework that continuously refines the model as new data are gathered. Safe and cost-aware optimization methods further support autonomous operation by reducing the risk of hardware damage and operational disruption.

\paragraph{Optimization of beam operation and accelerator design.}
Optimizing beam operation and accelerator component design presents significant computational challenges across two regimes: offline optimization, which prioritizes computational efficiency, and online optimization, which demands real-time decision-making and efficient use of facility time. Existing ML approaches include reinforcement learning, recurrent neural networks, hybrid genetic algorithm-neural network approaches, and advanced Bayesian optimization. Robustness and safety constraints can be embedded directly into the optimization process. 

A cornerstone of these strategies is the use of neural network surrogate models, which replace expensive conventional simulations during parameter searches and enable faster exploration of the solution space.  Maintaining optimized performance over time is as important as achieving it. Accelerator conditions drift continuously due to thermal effects, component aging, and environmental changes, causing performance degradation that requires frequent re-tuning. ML-based adaptive control algorithms — such as online reinforcement learning and continual learning approaches — that can track and compensate for such drift autonomously are an important and underexplored research direction.

\paragraph{Anomaly detection and monitoring.}
Anomaly detection and monitoring of beam incidents and detector operation are critical for maintaining safe and efficient accelerator operation. Bayesian regression is useful because it models probability distributions and provides uncertainty estimates that can define confidence intervals for anomaly detection after proper calibration. Dimensionality reduction techniques such as principal component analysis (PCA) followed by support vector machines (SVM), as well as convolutional neural network-based autoencoders, provide complementary approaches by compressing high-dimensional sensor data into lower-dimensional representations useful for different downstream tasks.

Depending on data availability, these approaches can be supervised, using labeled anomaly events, or unsupervised when such examples are scarce or unavailable. The main risks are uncontrolled false positives, which can produce false alarms  and erode operator trust, and false negatives, where critical anomalies go undetected. Both risks require careful threshold calibration and meaningful human oversight. Interpretability methods are also needed, particularly for unsupervised methods, to aid operators in understanding and responding to faults.

\paragraph{Virtual diagnostics.}
Virtual diagnostics provide a way to recover high-resolution information from limited measurements, such as reconstructing x-ray pulse characteristics that cannot be measured directly at the required fidelity or rate. Generative models, including diffusion models, can learn to produce high-dimensional outputs from low-dimensional observations, effectively performing reconstruction or super-resolution from sparse diagnostic data.

A practical workflow begins with high-dimensional simulation data used to train the mapping between limited and full diagnostic representations, followed by validation against real experimental data. The key risk is that models trained predominantly on simulation may produce outputs that are physically implausible or unreliable in operation. This can be mitigated by incorporating realistic measurement errors and noise directly into Monte Carlo training samples and by validating against real data whenever possible.

\paragraph{Real-time triggering and fast decision-making.}
Triggering systems represent one of the most demanding real-time computing challenges in collider physics, where decisions about which events to retain or discard must be made in microseconds. Neural networks implemented on optimized hardware-specific hardware, such as field-programmable gate arrays (FPGAs), can enable low-latency inference while keeping power and resource consumption within tight constraints. More advanced architectures, including transformers, may improve vertex reconstruction by exploiting correlations between tracks and energy deposits.

Deploying such models requires reducing model size through quantization and translating algorithms into firmware, often using high-level synthesis (HLS) tools. The major risks are difficult debugging, changing accelerator conditions, training bias, and limited interpretability. Because rejected events cannot be recovered, trigger models require rigorous validation, transparent development practices, and close collaboration between physicists and hardware engineers.

\paragraph{Human-AI interfaces.} For all of these challenges, the impact is limited by the ability of physicists and operators to integrate the AI methods into their daily workflows. AI reasoning models provide a potential solution via natural-language interfaces to users of lower-level AI/ML tools. Such systems can lower the cost of entry, e.g. allowing an operator to quickly deploy a new optimization tool, as well as expand the reach of AI/ML tools, e.g. through AutoResearch loops.


\subsection{Cosmology and Astrophysics}

A coherent AI/ML program for cosmology and astrophysics should focus on methods that improve information extraction while preserving physical control. The main solution areas are simulation-based inference, fast forward modeling, calibration and uncertainty propagation, anomaly detection, and reproducible shared infrastructure.

\paragraph{Simulation-based and field-level inference.}
Simulation-based inference and field-level inference provide a way to move beyond hand-designed summary statistics while retaining more information from maps, fields, catalogs, and joint data products. Learned likelihoods, posteriors, ratio estimators, neural compression, and reconstruction methods can relax restrictions imposed by traditional likelihood and covariance modeling, especially for higher-order statistics and combined probes.

The key requirement is interpretability and validation. These methods should reveal which scales, observables, and nuisance directions drive the result, and they must be tested for robustness against model misspecification and out-of-distribution data. Training and testing on different simulations, recalibrating posteriors, and comparing against conventional analyses are important mitigation strategies.

\paragraph{Fast forward modeling and surrogate simulation.}
Fast, accurate, and differentiable forward modeling is a major opportunity for cosmology and astrophysics. Emulators and hybrid surrogate models can accelerate nonlinear structure formation, baryonic effects, galaxy--halo modeling, instrument response, mock-catalog generation, and observable prediction. Generative models can also emulate mocks and observables, or paint small-scale physics into lower-resolution simulations.

The goal is not simply faster simulation, but forward models that are accurate where inference is sensitive and efficient enough for calibration, design, and posterior exploration at survey scale. The main risks are poor generalization across cosmologies, environments, simulations, or observing strategies, and over-correction that could accidentally remove new physics. These risks require careful benchmarks, uncertainty estimates, and comparison across simulation suites.

\paragraph{Compression, calibration, and uncertainty propagation.}
Neural networks can be used to cluster, compress, or summarize large data sets, ideally optimized against objectives such as mutual information. They can also support calibration, simulation-to-data adaptation, and scalable uncertainty propagation through multi-probe pipelines. Semi-supervised and self-supervised learning on survey data may help identify pathologies, selection effects, and mismatches between simulations and observations.

The key requirement is generalization. Methods must transfer across instruments, observing strategies, and simulation suites without silently absorbing mismatch into biased inference. Robustness checks, nuisance variation, recalibration, and explicit treatment of domain shift are therefore central parts of the solution.

\paragraph{Anomaly detection and discovery.}
Anomaly detection can support both discovery and systematic-error control in complex astrophysical data. Relevant applications include rare-event search, unusual lens or transient identification, structured contamination detection, and triage tools that prioritize follow-up. The goal should not be opaque anomaly scores alone, but calibrated decision support that separates genuine astrophysical novelty from instrument, simulation, or pipeline failure.

\paragraph{Reproducibility and shared infrastructure.}
Reproducibility requires more than publishing a final result. Useful AI/ML infrastructure should include code notebooks that run on manageable subsets of data, shared data and model hubs, curated catalogs, cross-matching tools, and benchmarks organized around surveys and data sets. Community platforms, including Hugging-Face-like resources, can support model sharing, validation, and reuse.

The practical challenges are funding, credit attribution, data quality, data variety, and community acceptance. Shared infrastructure is therefore both a technical and sociological solution: it can reduce duplicated effort, improve reproducibility, and make advanced AI/ML methods more accessible to the broader cosmology community.


\subsection{Neutrino Physics}

AI/ML solutions for neutrino physics should address reconstruction, simulation, rare-event detection, experimental design, and collaboration-scale infrastructure. The common goal is to improve physics sensitivity while reducing systematic uncertainties, computational cost, and duplication of effort across experiments.

\paragraph{Optimal event and particle reconstruction.}
Current ML approaches already vary across detector technologies. LArTPC reconstruction uses neural networks, water Cherenkov detectors such as Super-Kamiokande and Hyper-Kamiokande use graph neural networks (GNNs), Convolution Neural Networks (CNNs) on unrolled cylindrical detector representations, and differentiable models, while IceCube uses GNN-based approaches and hybrid ML plus maximum-likelihood methods for some event classes. These models demonstrate the promise of ML, but current approaches remain detector-specific and often rely on condensed information.

A major opportunity is to develop scalable reconstruction models that work closer to the hit level while producing particle-level and interaction-level outputs. This suggests a hierarchical reconstruction pipeline in which one model extracts features from detector hits, a second identifies particle instances, and a third reconstructs interaction-level information. Such a pipeline could support ring or cascade counting, segmentation, and reconstruction across wide energy ranges and detector technologies.

\paragraph{Training strategy and uncertainty.}
Building such reconstruction systems requires simulation data, hierarchical labels, and, where possible, small amounts of real or hand-labeled data for fine tuning. Models may be trained in stages and then tuned jointly, with possible use of unsupervised or self-supervised pretraining. A key risk is that fine tuning on limited calibration data can introduce bias outside the calibration region. Uncertainty estimation, interpretability, and validation against conventional likelihood-based reconstruction are therefore essential.

\paragraph{Accurate and unified simulation.}
Surrogate models and differentiable simulation provide a path toward faster and more flexible neutrino simulations. Possible targets include neutrino--nucleus cross-section models, particle propagation, optical photon propagation, detector response, and waveform generation. Surrogates can execute faster than conventional simulation, can be more compact than large lookup tables, and can be differentiable so that models can be tuned directly with data.

The main steps are to identify appropriate components for surrogate modeling, gather relevant training and calibration data, and validate transferability across detectors and physics regimes. Useful data sources include neutrino and electron scattering data, cosmic muons, laser calibration sources, and test beam data where available. The major risks are inadequate model choices, insufficient calibration data, insufficient computational resources, and long development timelines.

\paragraph{Anomaly detection for rare processes.}
Autoencoders, variational models, normalizing-flow-like models, and direct classifiers can support anomaly detection for rare processes. Autoencoders learn the structure of common backgrounds and define anomaly scores from reconstruction loss, while direct classifiers can be trained using simulated signal injected into noise or background data.

The main difficulty is that rare processes may not be present in the training data, and fake or simulated signals can introduce bias. Anomaly detection methods must therefore be validated carefully, with attention to whether they are used offline, online, or as part of a dedicated trigger. They should be sensitive to rare behavior while robust against detector noise, background fluctuations, and simulation artifacts.

\paragraph{Experimental design, construction, and operation.}
AI/ML can also support optimal experimental design, construction, and operation. Large language models can mine archives of past documents and experiments, support simulation code generation, assist with documentation, and provide interactive support for users, shifters, construction, and operations. Human-support hardware, including computer vision, cameras, and drones, could assist safety monitoring, installation, and maintenance.

For optimization tasks, differentiable simulation, surrogate models, and Bayesian optimization provide complementary tools. Gradient-based methods are useful when differentiable models are available; surrogate models are useful when resources are limited; and Bayesian optimization is useful when full computation is intractable. These tools can help engineering design at the subgroup level, integrate inputs from subgroups, and support execution of construction and operation.

\paragraph{Cross-experiment validation and shared baselines.}
Neutrino experiments differ in detector technology, data access policies, and analysis conventions, which makes cross-experiment validation difficult. A practical near-term goal is to establish shared benchmark tasks, common baseline models, and clear policies for what levels of data and simulation can be shared. These efforts would support reproducibility within neutrino physics while connecting naturally to the broader cross-domain infrastructure discussed in the following section.

\section{Cross-domain Research Opportunities}

Many challenges and opportunities for AI/ML research are shared across scientific fields. Using joint methods, shared benchmarks, and common infrastructure, interdisciplinary collaborations can be developed across cosmology and astrophysics, neutrino physics, and accelerator science. Through common effort, challenges such as the need for fast simulation, high-dimensional reconstruction, uncertainty propagation, domain shift mitigation, anomaly detection, real-time decision-making, and knowledge preservation can be addressed.
We outline below the main research topics where cross-pollination is expected to be most impactful. Some opportunities are common to all three fields, while others naturally connect two fields that face especially similar technical bottlenecks.

\subsection{Common Simulation, Surrogate Modeling, and Differentiable Pipelines}

Fast and accurate simulation is a central challenge across all three communities. Cosmology requires forward models that connect fundamental parameters to survey observables, including nonlinear structure formation, baryonic effects, and observational selection. Neutrino physics requires simulation of neutrino interactions, particle propagation, optical photon transport, detector response, and waveform generation. Accelerator physics requires fast models of beam dynamics, electromagnetic fields, component behavior, and facility operation.

A shared opportunity is the development of generative and surrogate models for fast simulation in a common framework. In cosmology, surrogates are needed to fill in small-scale physics in low-resolution gravity-only simulations, trained on more accurate but computationally expensive high-resolution runs. In neutrino physics, fast surrogates for photon propagation in particle detectors and for neutrino interaction models are actively being developed, with the latter potentially benefiting from models trained on electron scattering data. In accelerator physics, surrogate models are being explored for accelerator component modeling, virtual diagnostics, and inverse problems such as reconstructing phase-space distributions from limited measurements.

A modular approach, where individual simulation components can be replaced one at a time and benchmarked, offers a practical path toward more complete surrogate-based pipelines for complex experimental facilities. This approach is naturally shared across fields because all three communities face the same basic question: which parts of the simulation chain can be replaced by learned models without losing physical fidelity, interpretability, or uncertainty control?

Differentiable simulation provides similar cross-domain opportunities. By implementing a simulation code as a differentiable programming software---using e.g., JAX or Julia---gradients with respect to input parameters can be obtained simultaneously, enabling more efficient optimization and calibration compared to classical numerical methods. In high-dimensional inference problems, gradient and Hessian information can also improve sampling efficiency. In neutrino physics, differentiable implementations combining analytical models and neural network surrogates have been explored for detector response and event modeling. In accelerator physics, differentiable simulations with respect to parameters such as RF phase, magnet strength, and alignment errors can enable design optimization, sensitivity analysis, and operational tuning. In cosmology, differentiable forward models can support field-level inference and gradient-based exploration of large parameter spaces.

\subsection{High-Dimensional Reconstruction and Compact Representations}

Reconstruction is one of the most active areas of machine learning application across the fields, emerging in different forms. In cosmology, convolutional and graph neural networks are used to reconstruct dark matter distributions from galaxy images, infer photometric redshifts, and compress high-dimensional survey data. In neutrino physics, event reconstruction spans detector technologies and event topologies, from multi-ring events in water Cherenkov detectors to tracks and showers in LArTPCs and cascade morphologies in IceCube. In accelerator physics, virtual diagnostics reconstruct high-resolution beam or x-ray pulse properties from limited measurements.

The shared challenge is encoding high-dimensional, sparse, heterogeneous, and often non-Euclidean data into compact representations that preserve the information needed for inference, monitoring, and discovery. Sparse convolutional neural networks, graph neural networks, and transformer models have been used to address parts of this problem. Experience from one detector technology or data modality can therefore transfer naturally to others.

A particularly relevant common direction is the development of hierarchical representations. In neutrino physics, hierarchical datasets structured at the detector hit, particle, and interaction levels would enable more detailed reconstruction through segmentation and instance-level learning. Similar ideas appear in collider detectors, accelerator diagnostics, and survey analysis, where low-level measurements must be mapped to physically meaningful intermediate objects before entering final inference. Shared benchmarks and pretrained models on common challenges would significantly lower the barrier of entry for new experiments and researchers.

\subsection{Inference, Optimization, and Efficient Parameter Exploration}

Efficient exploration of large parameter spaces is a shared problem across all three communities. Simulators often involve many physics, detector, nuisance, and operational parameters, and brute-force exploration can be computationally prohibitive. In cosmology, a key difficulty is balancing multiple simultaneous objectives, such as reproducing galaxy clustering and stellar mass functions, while also modeling baryonic physics and survey systematics. In neutrino physics, fits can involve many detector, interaction, and systematic parameters. In accelerator physics, both design and operation require optimization over high-dimensional control spaces under safety and performance constraints.

Several solution strategies span multiple domains. Simulation-based inference, neural compression, Bayesian optimization and related methods (Bayesian algorithm execution, Bayesian optimal experimental design), reinforcement learning, differentiable simulation, and surrogate modeling can all reduce the cost of parameter exploration. Parameter correlations and degeneracies should be explicitly studied so that theory uncertainties, detector effects, and operational variables can be decoupled where possible. Methods should handle high-dimensional inputs and outputs, complex objective functions, multi-fidelity modeling, uncertainty quantification, and safety-aware exploration.

Design and model optimization provide a direct bridge between research fields. Methods developed for physics inference can be repurposed for detector and experimental design. In neutrino physics, accelerator-based beam lines can be optimized over target geometry, horn position, and shape, with operational constraints such as thermal and mechanical stress included. Detector configurations and calibration source placement can similarly be optimized. In accelerator physics, fast surrogate models for cavity performance or beam transport can be combined with uncertainty-aware optimization strategies. In cosmology, survey strategy, follow-up prioritization, and simulation design can benefit from the same ideas.

\subsection{Simulation-to-Data Mismatch, Domain Shift, and Uncertainty Propagation}

The gap between simulation and real data is a fundamental challenge in all three communities, though it manifests differently in each field. In cosmology, the Universe cannot be rerun, making controlled validation difficult. In accelerator physics, models trained in simulation may not transfer reliably to real machine operation. In neutrino physics, detector response, neutrino interaction models, and calibration limitations create persistent simulation-to-data discrepancies.

This makes domain adaptation and uncertainty propagation important cross-domain priorities. Transfer learning, semi-supervised and unsupervised fine-tuning, generative models such as diffusion models and normalizing flows, and explicit nuisance-parameter modeling can all help reduce simulation-to-data discrepancies. However, these methods also introduce additional challenges: models may hide mismodeling, over-correct real signals, or fail outside the calibration region.

A shared goal is therefore not only to improve agreement between simulation and data, but to propagate uncertainty through the full analysis chain. This includes experimental uncertainties, detector response uncertainties, calibration uncertainties, theoretical modeling uncertainties, and correlations across reconstruction, simulation, and final inference. Progress in this area would directly support precision measurements in cosmology, neutrino physics, and accelerator-based experiments.

\subsection{Anomaly Detection, Monitoring, and Discovery}

Anomaly detection serves different but complementary purposes across the fields. In accelerator physics, autonomous identification of faults, root cause localization, and operator alerting are critical operational needs. Pre-alert systems based on trend prediction could anticipate failures before thresholds are crossed. In neutrino physics, anomaly detection can support rare-event searches, data quality monitoring, trigger improvement, and identification of unusual detector behavior. In cosmology and astronomy, CCNs can classify pixel-level contributions from multiple sources including galaxies, stars, atmospheric emission, aircraft tracks, cosmic rays, and other sources.

The shared challenge is that unexpected signals are not known in advance. Anomaly detection is therefore inherently ill-posed: methods must be sensitive enough to identify rare or unexpected behavior while robust enough to avoid false alarms, detector artifacts, pipeline failures, and spurious discoveries. This makes anomaly detection a natural area for cross-community tool development, especially around benchmark datasets, uncertainty-aware anomaly scores, and human-in-the-loop validation.

\subsection{Real-Time Processing, Control, and Edge AI}

Online optimization and real-time decision-making are areas where machine learning can have immediate operational impact. In accelerator physics, smart online optimization of beam parameters using Bayesian optimization, recurrent neural networks, reinforcement learning, or hybrid methods is needed to maximize performance and maintain beam quality continuously. In neutrino and collider detectors, trigger decisions can be improved with online reconstruction models that offer better performance and faster inference than current approaches. In astronomy and satellite missions, edge AI can support local decision-making when bandwidth, latency, or operational constraints limit centralized processing.

The shared challenge is that real-time systems must make reliable decisions under strict latency and hardware constraints. Trigger systems, detector control, beam feedback, and online monitoring all face the same risk: fast decisions can improve data quality, but mistakes may result in unrecoverable data, interrupt operations, or introduce hard-to-diagnose biases. Hardware-software co-design, model compression, quantization, FPGA or edge deployment, and robust validation methods are therefore cross-cutting needs and would benefit from shared experiences.

\subsection{Shared Infrastructure, Benchmarks, Knowledge Bases, and Agentic Workflows}

A recurring theme across communities is the lack of common data formats, sharing policies, benchmark datasets, metrics, and reusable software infrastructure. Establishing standard datasets for well-defined machine learning tasks, such as generative modeling, classification, reconstruction, anomaly detection, and simulation-based inference, would facilitate comparison across methods and communities.

Large language models combined with retrieval-augmented generation offer a complementary path toward preserving and accessing institutional knowledge. Across large collaborations, important information is often stored in heterogeneous and poorly indexed formats, including wiki pages, public mailing list archives, logbooks, source code, and non-text data such as HDF5 files. LLM+RAG systems could support natural-language search, documentation generation, shift summaries, metadata suggestions, and software support, including translation of legacy code into modern equivalents.

Nearly all methods described here could benefit from agentic control, whether to integrate AI/ML tools, lower the barrier to entry, support physics discovery, or as a tool for operations. While specific knowledge bases may differ between fields, co-scientist tools could be broadly useful across domains, and would benefit from shared investment due to the cost of development.

This infrastructure problem is both technical and sociological. Productive cross-domain collaboration requires policies on what data can and should be shared, how it should be formatted, how analyses of shared data should be attributed, and who receives scientific credit. Shared tools, benchmarks, model hubs, and knowledge bases would reduce duplication of effort, minimize ``reinventing-the-wheel,'' lower the barrier of entry for new researchers, and enable common development across cosmology, neutrino physics, and accelerator science.

\subsection{Two-Field Transfer Opportunities}

Several opportunities connect especially strongly between two communities.

\paragraph{Cosmology and neutrino physics.}
Both fields rely heavily on simulation-based inference, indirect measurements, systematic-error modeling, and domain adaptation. Neutrino physics has developed practical strategies for dealing with detector calibration, simulation-to-data discrepancy, and limited control samples, which may be valuable for robust cosmological inference. Conversely, cosmology has developed mature tools for large-scale statistical inference, neural compression, and survey-scale validation that may transfer to neutrino analyses.

\paragraph{Neutrino physics and accelerator science.}
Neutrino beam experiments naturally connect accelerator design, beamline optimization, detector simulation, and physics sensitivity. Shared developments in surrogate modeling, differentiable simulation, Bayesian optimization, and operational monitoring could support end-to-end optimization from beamline parameters to final physics reach.

\paragraph{Accelerator science and cosmology/astrophysics.}
Both fields face large-scale scheduling, optimization, and control problems. Accelerator facilities optimize beam parameters and machine operation, while cosmology and astronomy optimize survey strategy, observing schedules, and follow-up decisions. Methods for safe optimization, real-time control, and operational decision support may transfer between these domains, especially when combined with uncertainty-aware surrogate models.


\section{Summary}

This white paper identified major scientific challenges and AI/ML opportunities across accelerator physics, cosmology and astrophysics, and neutrino physics discussed during the FAIRS Japan 2024 event. While the scientific targets differ across these fields, the technical obstacles show strong commonality. Each community must extract reliable physical information from complex data, model systems that are expensive or incomplete to simulate, propagate uncertainties through long analysis chains, and make robust decisions under practical constraints.

For accelerator physics, the main opportunities lie in improving the design, optimization, monitoring, and operation of complex beam facilities. AI/ML methods can accelerate simulation, support online tuning, enable virtual diagnostics, detect anomalies, and assist real-time decision-making. These capabilities are directly connected to improved facility performance, reliability, safety, and scientific output. Surrogate models and digital twins that emulate accelerator components and beam dynamics represent an additional key opportunity, enabling fast optimization and autonomous operation.

For cosmology and astrophysics, the main opportunities lie in extracting maximal information from large observational surveys while controlling theoretical, observational, and astrophysical systematics. AI/ML methods can support simulation-based inference, neural compression, fast forward modeling, anomaly detection, and reproducible analysis infrastructure. These approaches are essential for next-generation surveys where systematic uncertainties increasingly dominate over statistical limitations.

For neutrino physics, the main opportunities lie in event reconstruction, accurate and unified simulation, rare-event detection, and experimental optimization. AI/ML methods can help reconstruct complex final states, reduce simulation cost, improve treatment of detector response and interaction modeling, identify rare signals, and optimize detector and beamline designs toward physics reach.

Across all domains, several research directions stand out as especially important. Fast surrogate models and differentiable simulation can reduce computational cost while enabling calibration, optimization, and inference. High-dimensional reconstruction methods can preserve more information from raw detector or survey data. Domain adaptation and uncertainty propagation are essential for making simulation-trained models reliable on real data. Anomaly detection can support both discovery and operational monitoring. Real-time processing and edge AI can improve triggers, feedback systems, and autonomous control.

Equally important are the social and infrastructural components of AI/ML research. Shared datasets, benchmarks, software frameworks, model hubs, documentation tools, and knowledge bases are necessary to make progress reproducible and reusable. These resources can reduce duplication across communities, support cross-domain transfer, and allow researchers to build on common developments rather than repeatedly solving similar problems in isolation.

The path forward should therefore combine domain-specific AI/ML applications with cross-domain coordination. Near-term progress can come from focused demonstrators, benchmark tasks, and modular tools that solve concrete problems within individual experiments. Long-term progress will require shared infrastructure, common validation practices, and sustained collaboration between domain scientists, AI/ML researchers, software experts, and computing facilities.

AI/ML will not replace physical modeling, experimental expertise, or careful statistical reasoning. Its greatest value lies in augmenting them: making simulations faster, data analyses more informative, operations more reliable, and collaborations more effective. If developed with focus on uncertainty, interpretability, reproducibility, and shared infrastructure, AI/ML can substantially enhance the scientific reach of future physics experiments.


\section*{Acknowledgments}
This workshop was supported by Center for Data-driven Discovery (CD3), Kavli Institute for Physics and Mathematics of the Universe (KIPMU) and Kobayashi-Maskawa Institute for the Origin of Particles and the Universe (KMI), Nagoya University. VM is supported by JST EXPERT-J, Japan Grant Number JPMJEX2509. LT is supported by JSPS under KAKENHI 24K22878 and 26K17136 and by the Royal Society under ICA\textbackslash R2\textbackslash 252140. NMN is supported by JSPS KAKENHI Grant Number 25K23373 and 26H00404. The R\&D works are supported by the research project at RCNP U. Osaka, the interdisciplinary co-creation project at D3C U. Osaka, and the joint research program at ICEPP U.Tokyo. The work is also supported by JSPS Grant-in-Aid for Transformative Research Areas A JP22H05113 and International Leading Research JP22K21347. AJN is supported by JSPS Kakenhi Grant numbers, JP25H01551 and JP22K21349. TT’s work is partially supported by the 34th (FY 2024) Academic research grant (Natural Science) No.~9284 from DAIKO FOUNDATION and by JSPS KAKENHI Grant Number 26H00403. HM is supported by JSPS Kakenhi Grant Numbers, JP23H00108 and JP24KK0065 and by JST FOREST Grant Number JPMJFR2129.

\newpage 
\bibliographystyle{unsrt}
\bibliography{references}
\end{document}